\documentclass[10pt]{article}
\usepackage[latin1]{inputenc}
\usepackage[english]{babel}
\usepackage[T1]{fontenc}
\usepackage{url}
\usepackage[left=1in,right=1in,top=1in,bottom=1in]{geometry}
\usepackage[numbers]{natbib}
\usepackage{graphicx}
\usepackage[draft]{minted}
\usepackage{authblk}

\title{Designing and building the mlpack open-source machine learning library}
\author[1]{Ryan R. Curtin\thanks{ryan@ratml.org}}
\author[2]{Marcus Edel\thanks{marcus.edel@fu-berlin.de}}
\affil[1]{Center for Advanced Machine Learning, Symantec Corporation}
\affil[2]{Institute of Computer Science, Free University of Berlin}
\date{August 18, 2017}

\begin{document}
\maketitle

\begin{abstract}
mlpack is an open-source C++ machine learning library with an emphasis on
speed and flexibility.  Since its original inception in 2007, it has grown to be
a large project implementing a wide variety of machine learning algorithms, from
standard techniques such as decision trees and logistic regression to
modern techniques such as deep neural networks as well as other
recently-published cutting-edge techniques not found in any other library.
mlpack is quite fast, with benchmarks showing mlpack outperforming other
libraries' implementations of the same methods.  mlpack has an active community,
with contributors from around the world---including some from PUST.  This short
paper describes the goals and design of mlpack, discusses how the open-source
community functions, and shows an example usage of mlpack for a simple data
science problem.
\end{abstract}

\section{Introduction}

The broad field of machine learning has seen a huge surge in popularity in the
past few decades, with compelling results for problems like image recognition
\cite{ciregan2012multi}, automatic medical diagnostics
\cite{kononenko2001machine}, recommender systems \cite{bell2007lessons}, and
speech recognition \cite{graves2013speech}.  Loosely speaking, any machine
learning algorithm depends on data, and the quality of the algorithm's results
can be increased by adding more data \cite{halevy2009unreasonable}.  Since mass
quantities of data are now so widely available, it is important to enable
practitioners to run machine learning algorithms on large datasets.  Thus, it is
important to provide high-quality implementations of these algorithms.

This reality has motivated us to design, build, and maintain the open-source
{\bf mlpack} machine learning library, which focuses on providing high-quality
implementations of both typical and cutting-edge machine learning algorithms to
machine learning practitioners around the world.  The library is free to use
under the terms of the BSD open-source license, and accepts contributions from
anyone.

mlpack development began in 2007 at the Georgia Institute of Technology, and the
first stable release of the library was in 2011 \cite{mlpack2013}.  Since then,
the project has grown to have nearly 100 individual contributors from around the
world, and at the time of this writing has been downloaded over 50,000 times and
used in close to 100 academic papers.  The growth and success of mlpack can be
attributed in part to its open community and its participation in open-source
initiatives such as Google Summer of Code
(\url{https://summerofcode.withgoogle.com/}).

This short document aims to discuss the design and usage of mlpack, as well as
how the open-source community functions.  Section \ref{sec:description} discusses the
design of mlpack, providing an overview of its functionality.  Section
\ref{sec:community} discusses the open-source community of mlpack and its
participation in programs such as Google Summer of Code, and how someone new to
the library might get involved and contribute.  Then, Section \ref{sec:design}
discusses the design and API considerations of mlpack.  Section
\ref{sec:example} shows a simple example of how mlpack can be used to solve a
typical data science problem.  Lastly, Section \ref{sec:future} briefly
discusses the future planned functionality and directions of mlpack.

\section{Description of mlpack}
\label{sec:description}

The mlpack library aims to balance four major goals:

\begin{itemize} \itemsep -2pt
  \item implement {\bf fast, scalable} machine learning algorithms;
  \item design an {\bf intuitive, consistent, and simple} API for non-expert
users;
  \item implement a {\bf variety} of machine learning methods; and
  \item provide {\bf cutting-edge} machine learning algorithms unavailable
elsewhere.
\end{itemize}

In order to provide efficient implementations, mlpack is written in the C++
language, and uses C++ template metaprogramming to optimize mathematical
expressions when the program is compiled.  In addition, mlpack uses the fast
Armadillo linear algebra library as its core \cite{sanderson2016armadillo}.

Inside of mlpack, a breadth of machine learning algorithms are available.  A
basic list is given below; but note that the list is not exhaustive, and to find
an up-to-date list one should consult the mlpack website.

\begin{itemize} \itemsep -2pt
  \item {\bf Classification}: logistic regression, softmax regression, deep
neural networks, AdaBoost, perceptrons, Naive Bayes classifier, decision trees,
decision stumps, Hoeffding trees
  \item {\bf Regression}: linear regression, ridge regression, LARS, LASSO
  \item {\bf Clustering}: $k$-means, DBSCAN, minimum spanning tree calculation
  \item {\bf Data transformations}: PCA, kernel PCA, sparse coding, local
coordinate coding, RADICAL, NMF
  \item {\bf Distance-based tasks}: (approximate and exact) nearest neighbor
search, furthest neighbor search, range search, max-kernel search, locality
sensitive hashing
  \item {\bf Optimization algorithms}: L-BFGS, SGD, gradient descent, simulated
annealing, AdaGrad, ADAM, RMSprop, AdaDelta, augmented Lagrangian solver
  \item {\bf Other}: collaborative filtering/recommender systems, data
preprocessing tools, neighborhood components analysis, SVD matrix
decompositions, density estimation trees
\end{itemize}

\section{Community description}
\label{sec:community}

mlpack is an open-source project whose development is primarily conducted on
Github (\url{https://github.com/mlpack/mlpack}) and on IRC (internet relay
chat).  Github is a centralized open-source development website that allows any
user to contribute easily to some open-source software project.  The website can
be used to do many things---a user can ask for help using mlpack, report a bug
inside of mlpack, submit new code, discuss ideas with mlpack maintainers, update
or add documentation, or run automatic tests on mlpack code, among other things.
Thanks to Github's standard interface and wide adoption inside the open-source
community, this means that it is quite simple and straightforward for anyone
interested in mlpack to perform any of these tasks.

\begin{figure}
\begin{center}
  \includegraphics[width=0.65\textwidth]{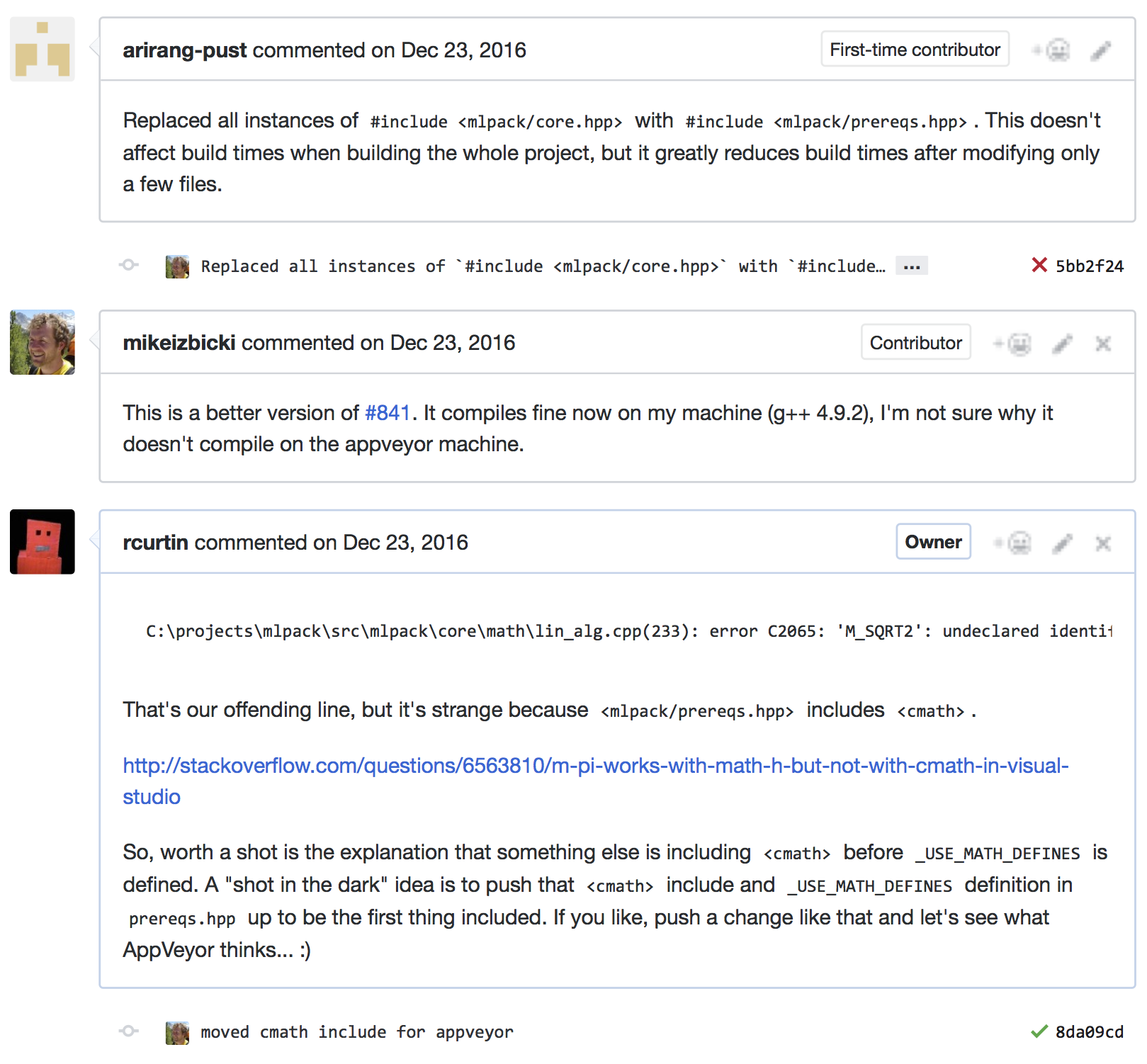}
\end{center}
\vspace*{-1.5em}
\caption{Typical pull request discussion for an mlpack contribution.}
\label{fig:pr}
\vspace*{-1em}
\end{figure}

When a user submits new code to mlpack for inclusion as a {\it pull request},
the code must be reviewed by mlpack maintainers to ensure that the code meets
the mlpack standards.  These standards include simple requirements like
formatting, and more complex standards like the standardized APIs described in
the following section.  In addition, any new code must be submitted with tests,
so that it can be known that the new code functions properly.  Typically the
review process will include several comments from maintainers that need to be
addressed by the contributor(s).  When agreement is reached between the
maintainers and the contributors that the code is ready, then the code can be
merged into the mlpack codebase.  Since the mlpack project is careful to accept
only high-quality code, this process can typically take several days to several
weeks, depending on the complexity and quality of the original submission.  An
example of this process can be seen at
\url{https://github.com/mlpack/mlpack/pull/842}.  A screenshot of some of the
discussion is pictured in Figure \ref{fig:pr}.

Because the Github pull request interface and process is standardized across all
Github projects, it is straightforward for new contributors to get involved with
mlpack and start getting their contributions merged into the code.  Since the
mlpack project moved to Github in the fall of 2014, the number of contributors
to the project has grown more rapidly than before.

Another factor contributing to the growth of the project is mlpack's
participation in the Google Summer of Code program
(\url{https://summerofcode.withgoogle.com}).  The Google Summer of Code program
is an open-source initiative led by Google, with the goal of getting more
students to participate in the development of open-source software.  In Summer
of Code, the mlpack library selects a number of promising students for
acceptance into the program.  These students each prepare a proposal for a
project that they will complete over the course of the summer, with a detailed
description and timeline.  Those students who are selected are paid a stipend
and a t-shirt by Google to work full-time on their project over the summer, and
usually at the end of the summer their work will be incorporated into the
library.  Communication during the summer is done usually via real-time IRC or
on Github, and the use of real-time chat allows students and mentors to build
relationships, which in turn strengthens the community.

Examples of previous improvements include an automatic benchmarking system
\cite{edel2014automatic}, a collaborative filtering framework
\cite{agrawal2015collaborative}, implementation of different tree types
\cite{mcnames2001fast}, addition of new deep learning techniques
\cite{andrychowicz2016learning}, and others.  Since 2013, mlpack has mentored 24
students from around the world.

\section{Design of mlpack}
\label{sec:design}

An important consideration of any library is the interface it implements (the
API).  If the interface is confusing, then it will be difficult for end users to
solve their problems.  Therefore, in the design of mlpack we have endeavored to
make each piece of the library match a consistent and intuitive interface.  This
is similar to the efforts of the Python open-source machine learning library
{\tt scikit-learn} \cite{buitinck2013api}.

The top-level interface to mlpack is its machine learning algorithms, listed
earlier in Section \ref{sec:description}.  Each of these classes of algorithms
is typically used to solve a different task.  To take an example, for a
classification algorithm, given a data point, we wish to
select a distinct class that the data point belongs to.  An example in the field
of malware detection would be to classify a file (which is a data point) as {\tt
malicious} or {\tt benign}.

Inside of mlpack, each class of algorithm must implement an identical API.  This
allows easy combination and substitution of mlpack algorithms: a common
requirement in data science tasks is to try multiple different algorithms to
solve a single problem.  If each of these algorithms implements the same API,
then the developer can simply switch the type of algorithm used without needing
to adapt their support code.

All classification algorithms inside of mlpack, such as the Naive Bayes
classifier or the decision tree, must implement the same interface for two
methods: {\tt Train()} and {\tt Classify()}.  The {\tt Train()} function takes
training data and labels and trains the classifier on the data.  The {\tt
Classify()} function then can use this trained model to predict classes for the
given test points.  For instance, the code below could be used to train an
mlpack classifier on the matrix {\tt dataset} with labels {\tt labels}, and then
predict the class of the test point {\tt (1.3, 2.4, 1.0)}.

{\small
\begin{minted}{c++}
        // Here we will use the DecisionTree class, but we could easily use another
        // mlpack classifier.
        typedef DecisionTree<> ClassifierType;

        // Train the model.
        ClassifierType classifier;
        classifier.Train(dataset, labels);

        // Now predict on the test point.
        arma::vec point("1.3, 2.4, 1.0");
        const size_t predictedClass = classifier.Classify(point);

        std::cout << "Predicted class is " << predictedClass << "." << std::endl;
\end{minted}
}

In addition to this top-level interface, lower-level interfaces inside mlpack
are also required to implement identical APIs for easy composability.  Examples
of this include the {\tt TreeType} API, which provides a generic interface for
space partitioning trees such as the $kd$-tree \cite{curtin2013tree}, the {\tt
OptimizerType} API, which provides a simple interface for numerical optimization
tasks, and the {\tt KernelType} API, which provides a simple interface for
machine learning kernel functions \cite{hofmann2008kernel}.

The description of the mlpack API given here is certainly not comprehensive; we
would refer readers to the documentation found at
\url{http://www.mlpack.org/docs.html} for more information.

Another important component of mlpack design is flexibility.  To this end, we
use a C++ programming paradigm known as {\it policy-based design}
\cite{alexandrescu2001modern}.  In short, this means that the behavior of mlpack
algorithms is easily controllable by the user, simply by specifying template
parameters.  For example, the {\tt DualTreeBoruvka} class, which is used to
calculate minimum spanning trees \cite{march2010euclidean}, accepts a template
parameter {\tt MetricType}.  Thus, if the user wants to find a minimum spanning
tree in a metric space that mlpack does not have support for, they merely need
to implement a distance metric class, and then they can use the type {\tt
DualTreeBoruvka<MyCustomMetric>} to calculate the minimum spanning tree in the
custom metric space.  Additionally, because templates are being used, there is
no additional overhead, as there would be when providing this type of support
through inheritance or in other languages such as Python or C.

\section{Example data science application}
\label{sec:example}

This section details the usage of mlpack for a simple data science task.  In
this example, we would like to predict the primary type of tree in a forest
based only on cartographic variables such as elevation, slope, azimuth, soil
type, and others.  The dataset we use for this task is known as the {\tt
covertype} dataset \cite{blackard1999comparative}, and is available on the UCI
machine learning dataset repository \cite{uci}.  The version of mlpack used in
this example is mlpack 2.2.3, released on May 24, 2017.

In this example, we will show how we can use mlpack's decision tree to build a
classifier for this task, and we will obtain the accuracy of the classifier on a
held-out test dataset.

First, we would like to load the dataset and labels.  Suppose these are in the
files {\tt covertype.csv} and {\tt covertype.labels.csv}, respectively.  We will
load the dataset into an Armadillo matrix called {\tt dataset}, and we will
load the labels into an unsigned Armadillo matrix called {\tt labels}.  This can
be done with mlpack's {\tt Load()} function, in the code shown below.

{\small
\begin{minted}{c++}
        Load("covertype.csv", dataset);
        Load("covertype.labels.csv", labels);
\end{minted}
}

\noindent Next we need to split the dataset into a training set (only used for the
training process) and a test set (which is what we will use to obtain the
accuracy of the classifier).  We can do this using the {\tt
Split()} function, setting aside 20\% of the data for the test set:

{\small
\begin{minted}{c++}
        // Split dataset/labels into trainingData/trainingLabels and
        // testData/testLabels.  The 'labels' object needs to be cast into an
        // arma::Row<size_t> object.
        Split(dataset, arma::Row<size_t>(labels.row(0)), trainingData,
            testData, trainingLabels, testLabels, 0.2);
\end{minted}
}

\noindent Now, we can use the training data and labels to train a decision tree
classifier.

{\small
\begin{minted}{c++}
        // Optionally, the constructor can be used to set specific algorithm
        // parameters.  We will leave them as the defaults here.  7 is the
        // number of classes in the dataset.
        DecisionTree<> treeClassifier;
        treeClassifier.Train(trainingData, trainingLabels, 7);
\end{minted}
}

\noindent The training may take some time.  Following training, we want to obtain
predictions for the test set.  We will store the predictions of the decision
tree in the variable {\tt predictions}.

{\small
\begin{minted}{c++}
        // Predict the classes of the points in testData and store them in the output
        // predictions vector.
        arma::Row<size_t> predictions;
        treeClassifier.Classify(testData, predictions);
\end{minted}
}

\noindent Lastly, we would like to calculate the accuracy of each classifier.  We can do
this by using an Armadillo expression that will sum the number of identical
elements in the test labels and the prediction vectors.

{\small
\begin{minted}{c++}
        // Calculate the number of points that the classifier got right.
        const size_t correct = arma::accu(predictions == testLabels);

        // Now print the accuracy.
        const double accuracy = double(correct) / double(testLabels.n_elem);
        std::cout << "Decision tree classifier accuracy on test set: "
            << (accuracy * 100.0) << "." << std::endl;
\end{minted}
}

\noindent Using this code, we can compile the program\footnote{Full sources and
compilation instructions given in the appendix.}, and when we run it we receive
the following output:

{\small
\begin{verbatim}
        Decision tree classifier accuracy on test set: 71.0022.
\end{verbatim}
}

So we can see that the decision tree classifier can obtain reasonable accuracy
on our data.  If we were to tune the parameters of the classifier, we could
likely achieve better accuracy.  Also, if we desired more predictions for new
test points, as in a production machine learning application, we could easily
use the {\tt Classify()} method.  This simple example, of course, does not
discuss all the intricacies of good data science work, but instead presents how
mlpack could be used in a simple data science context.  Adapting the example to
a real-life application would be more complex but straightforward.

\section{Conclusion and Future Goals}
\label{sec:future}

We have introduced the open-source mlpack C++ machine learning library and
discussed its general design philosophy, as well as its open-source development
process.  However, mlpack is under active development continually, so the
library is far from complete or static.  New versions are released typically
every few months with enhanced functionality, bugfixes, and efficiency
improvements.  The next major release, mlpack 3.0.0, is currently planned for
late in 2017 with the following major functionality improvements and changes:

\begin{itemize} \itemsep -2pt
  \item Bindings to Python and other languages
  \item Addition of random forests
  \item Addition of deep neural network toolkit
  \item Addition of several optimizers and changes to optimizer API
  \item Cross-validation and hyper-parameter tuning framework
\end{itemize}

This list, of course, is non-comprehensive; for a more clear list of goals, the
mlpack website (\url{http://www.mlpack.org/}) and the Github website
(\url{https://github.com/mlpack/mlpack}) have more details both about the future
goals of the project and the current status of the project.

\bibliographystyle{plain}
\bibliography{paper}

\newpage
\appendix
\section{Data science example code}

The full code for the example in Section \ref{sec:example} is given below.

{\footnotesize
\begin{minted}{c++}
/**
 * Build a decision tree classifier on the covertype dataset, then print the
 * accuracy of the classifier on the held-out testing data.
 */
#include <mlpack/core.hpp>
#include <mlpack/core/data/split_data.hpp>
#include <mlpack/core/data/load_impl.hpp>
#include <mlpack/methods/decision_tree/decision_tree.hpp>

using namespace mlpack;
using namespace mlpack::data;
using namespace mlpack::tree;

int main()
{
  // Load the dataset.
  arma::mat dataset;
  arma::Mat<size_t> labels;

  Load("covertype.csv", dataset);
  Load("covertype.labels.csv", labels);

  // Split into training and test dataset.
  arma::mat trainingData, testData;
  arma::Row<size_t> trainingLabels, testLabels;

  // Split dataset/labels into trainingData/trainingLabels and
  // testData/testLabels.
  Split(dataset, arma::Row<size_t>(labels.row(0)), trainingData, testData,
      trainingLabels, testLabels, 0.2);

  // Optionally, the constructor can be used to set specific algorithm
  // parameters.  We will leave them as the defaults here.  7 is the
  // number of classes in the dataset.
  DecisionTree<> treeClassifier;
  treeClassifier.Train(trainingData, trainingLabels, 7);

  // Predict the classes of the points in testData and store them in the output
  // predictions vector.
  arma::Row<size_t> predictions;
  treeClassifier.Classify(testData, predictions);

  // Calculate the number of points that the classifier got right.
  const size_t correct = arma::accu(predictions == testLabels);

  // Now print the accuracy.
  const double accuracy = double(correct) / double(testLabels.n_elem);
  std::cout << "Decision tree classifier accuracy on test set: "
      << (accuracy * 100.0) << "." << std::endl;
}
\end{minted}
}

Assuming that mlpack is installed on the system, then this code ({\tt code.cpp})
can be compiled with {\tt g++} or another compiler:

\begin{verbatim}
$ g++ -o code code.cpp -std=c++11 -lmlpack -larmadillo
\end{verbatim}

Typically, installing mlpack is easy, especially if a Linux distribution such as
Ubuntu, Debian, or Fedora is being used.  Installing mlpack on an Ubuntu system
can be done with the following command:

\begin{verbatim}
$ sudo apt-get install libmlpack-dev
\end{verbatim}

\end{document}